\documentstyle[prl,aps,epsf,multicol]{revtex}

\begin{document}

\title{Crossover from Isotropic to Directed Percolation}

\author{Per Fr\"ojdh and Marcel den Nijs}

\address{Department of Physics, University of Washington,
 P.O. Box 351560, Seattle, Washington 98195-1560}

\maketitle
\begin{abstract}
Directed percolation is one of the generic universality classes for 
dynamic processes. We study the crossover from isotropic to directed 
percolation by representing the combined problem as a random cluster 
model, with a parameter $r$ controlling the spontaneous birth of new forest 
fires. We obtain the exact crossover exponent $y_{DP}=y_T-1$ at $r=1$
using Coulomb gas methods in 2D. Isotropic percolation is stable, as is
confirmed by numerical finite-size scaling results. For $D \geq 3$,
the stability seems to change. An intuitive argument, however, suggests that
directed percolation at $r=0$ is unstable and that the scaling properties 
of forest fires at intermediate values of $r$ are in the same universality 
class as isotropic percolation, not only in 2D, but in all dimensions.
\end{abstract}

\begin{multicols}{2}
\narrowtext

Directed percolation (DP) has emerged during recent years as one of the
most common dynamic universality classes. It applies to a wide array of 
dynamic processes, ranging from flow through a porous medium in a 
gravitational field, forest fires and epidemic growth, to surface chemical 
reactions \cite{DPrev}.
In $1+1$ dimensions the DP critical exponents are known accurately from 
numerical studies in the early eighties \cite{Kinzel}. Today the numbers 
have been refined \cite{Jensen}, but analytic insight is still lacking. 
The ultimate goal is to understand scale invariance in DP at the same 
level as that of isotropic percolation (IP). This will require some sort 
of generalization of conformal invariance \cite{confinv} and the Coulomb 
gas method \cite{Nienhuis,Marcel}, which apply to isotropic scaling 
phenomena in $1+1$ dimensions. The scaling properties of dynamic processes 
like DP are intrinsically anisotropic. Studies of crossover phenomena, 
like the one presented here, are a start into this direction.

Consider a square lattice, as shown in Fig.~1. For IP the bond between 
vertices $i$ and $j$ represents a channel that can be open or closed, 
$s_{ij}=\pm 1$, with a probability $p$ or $1-p$, respectively.
At $p_c= \frac{1}{2}$, an infinite cluster of connected channels appears, 
such that fluid can percolate all the way through the lattice from one 
end to the other. The correlation length diverges as 
$\xi \sim |p-p_c|^{-\nu}$ with $1/\nu=y_p=y_T=3/4$ \cite{Marcel}.

\begin{figure}
\centerline{\vbox{\epsfxsize=65mm \epsfbox{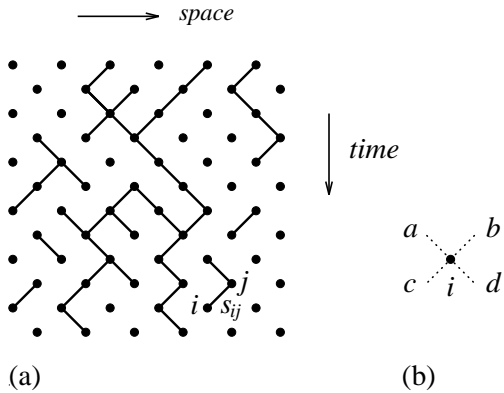}}}
\vspace{2ex}
\caption{Bond percolation on a square lattice with a time-like vertical
direction: (a) Open bonds ($s_{ij} = 1$) are marked with solid lines.
(b) Bonds around each vertex $i$ are labeled $s_a$, $s_b$, $s_c$, and
$s_d$.}
\end{figure}

Directed percolation describes the same type of flow in a 
porous medium, but in the presence of a gravitational field. 
The fluid can only flow downwards (positive time direction in Fig.~1). 
Points at either end are considered connected only if there exists 
a path between them without the need to back-track in time.
The DP threshold is larger, $p_c=0.6447$, and the scaling properties 
at $p_c$ are anisotropic. The time-like ($\parallel$) and spatial 
($\perp$) correlation lengths diverge with different exponents, 
$\xi_\parallel \sim |p-p_c|^{-\nu_\parallel}$ and 
$\xi_\perp \sim |p-p_c|^{-\nu_\perp}$, respectively, with 
$z=\nu_\parallel / \nu_\perp = 1.581$ the dynamic critical exponent and
$y_p = 1/\nu_\perp = 0.9117$ the scaling dimension of $p - p_c$. 
The values of these exponents are known only numerically \cite{Jensen}.

One of the alternate incarnations of DP is as a model for the propagation
in time of a forest fire on a $(D-1)$-dimensional lattice.
DP treats burnt trees as equivalent to never-burnt trees.
This absence of immunization is unrealistic, unless we visualize this as a 
fire inside a $D$-dimensional forest in a strong wind. 
The fire propagates only in the 
direction of the wind, which maps out time. The difference between IP and 
DP is the possibility to spontaneously ignite new forest fires. 
Define two operators (see Fig.~1):
\begin{equation}
\label{C_op}
{\cal C}_i = \frac{1}{4} (1-s_a)(1-s_b)
\left[ 1 - \frac{1}{4} (1-s_c)(1-s_d) \right] 
\end{equation}
equals one if a new forest fire is created at vertex $i$ and is 
zero otherwise.
\begin{equation}
\label{A_op}
{\cal A}_i = \frac{1}{16} (1-s_a)(1-s_b)(1-s_c)(1-s_d) 
\end{equation}
equals one if vertex $i$ is not part of any forest fire
and is zero otherwise. Consider the following partition function
\begin{equation}
\label{partition_function}
{\cal Z} = \sum_{\cal G} p^{N_b} (1-p)^{2N_v-N_b} q^{N_c} 
\prod_{i} r^{{\cal C}_i} \gamma^{{\cal A}_i} ,
\end{equation}
where $N_v$ is the number of vertices of the lattice,
$N_b$ the number of percolating bonds in graph ${\cal G}$, and
$N_c$ the number of disconnected clusters in ${\cal G}$ (including the ones
that contain only one vertex). $0 \leq r \leq 1$ controls the spontaneous 
birth of forest fires, and the factors
\begin{equation}
\gamma = \frac{1-rp(2-p)}{(1-p)^2}
\end{equation}
ensure that the partition function remains stochastic 
at $q=1$ for all values of $r$.
At $r=1$ we recover the conventional random cluster (RC) model \cite{FK},
which is equivalent to the $q$-state Potts model and describes
IP in the $q=1$ limit \cite{Wu}.

At $r=1$ the RC model can be mapped onto the 6-vertex model
and in the continuum limit onto a sine-Gordon type model. The latter can be 
rephrased as a Coulomb gas between charged plates \cite{Marcel}.
These equivalences lead in the early eighties to the derivation of the 
exact values of the critical exponents of the Potts model 
\cite{Nienhuis,Marcel}. 
The same techniques can be applied to determine the stability of IP 
with respect to $r$.
At $r=1$ and $p=\frac{1}{2}$ the crossover operator takes the form
\begin{eqnarray}
{\cal O}_i & = & \frac{1}{8} (s_a + s_b + s_c + s_d)+ 
  \frac{1}{8} (s_c + s_d - s_a - s_b) \nonumber \\
& - & \frac{1}{4} \left[ s_c s_d + 2(s_a + s_b)(s_c + s_d) \right] + O(s^3).
\label{DP_op}
\end{eqnarray}
First we need to substract from Eq.~(5) operators that are already
present in the theory at $p_c$ and $r=1$. Those operators are invariant 
under time-reversal $T$ ($s_a \leftrightarrow s_c$ and 
$s_b \leftrightarrow s_d$; see Fig.~1).
For example, the spin variables $s_{ij}$ represent the energy density
of the Potts model. It is invariant under $T$ and changes sign 
under the duality transformation ($s_{ij}\to -s_{ij}$).
The first term on the right hand side of Eq.~(\ref{DP_op}) 
is the energy density. Its presence indicates that the critical line 
approaches $r=1$ with a finite angle.
The second term has the form of the gradient of the energy operator 
with respect to time, and breaks time reversal.
This is the true crossover operator.
It is essential not to miss any topological operators that might be
hiding in ${\cal O}_i$.
This is where the mapping to the Coulomb gas becomes important.
We have determined the Coulomb gas representation of the
crossover-operator pair-correlation function.
The details are not presented here, since
we find no hidden topological operators. 
The most dominant component in ${\cal O}_i$ 
that breaks time reversal is simply
the gradient of the energy operator with respect to time,
\begin{equation}
\label{Cross_op}
{\cal O}_i \sim {\partial \epsilon \over\partial t}
\end{equation}
Hence, the crossover exponent is equal to $y_{DP}=y_T-1$.
This is an exact result. The crossover exponent is irrelevant, 
$y_{DP}=-1/4$ is negative since $y_T=3/4$.
This indicates that the scaling properties of a single forest fire, 
DP at $r=0$, are different from those of forests in which new fires can 
spontaneously ignite, with probability proportional to $r$. 
Those must have the same scaling properties as IP. 
The following numerical results confirm this.

We have performed a finite-size scaling (FFS) analysis of 
Eq.~(\ref{partition_function}) at $q=1$. The transfer matrix acts on a 
state vector which in its most efficient representation contains only 
slightly less than $5^N$ elements for system size $N$. 
Non-local connectivity information is required at each moment in time.
It does not suffice to know which trees are on fire and which are not.
We need to know which burning trees at time $t$ are part of the same cluster
via a path in the past. Our coding for this is inspired by an earlier 
FSS analysis at $r=1$ by Bl\"ote and Nightingale \cite{BN}.
The full connectivity can be reconstructed by
attributing to each vertex 5 possible states: 
$(0,\Leftarrow,\Rightarrow, \Leftrightarrow, \star)$.
Vertex $i$ is not part of any forest fire (state 0) if both trees just 
above it are not burning, $s_a = s_b= -1$ (see Fig.~1), but is
part of the fire if either or both are burning. 
In state $\Leftarrow$  at least one vertex to the left of $i$
is part of the same cluster, but none to its right. 
In state $\Rightarrow$ at least one vertex to the right
is part of the same cluster, but none to its left.
In state $\Leftrightarrow$ at least two vertices, one on each side,
is part of the same cluster.
In state $\star$ no other vertices at this moment in time
are part of the same cluster. 

\begin{figure}
\centerline{\vbox{\epsfxsize=80mm \epsfbox{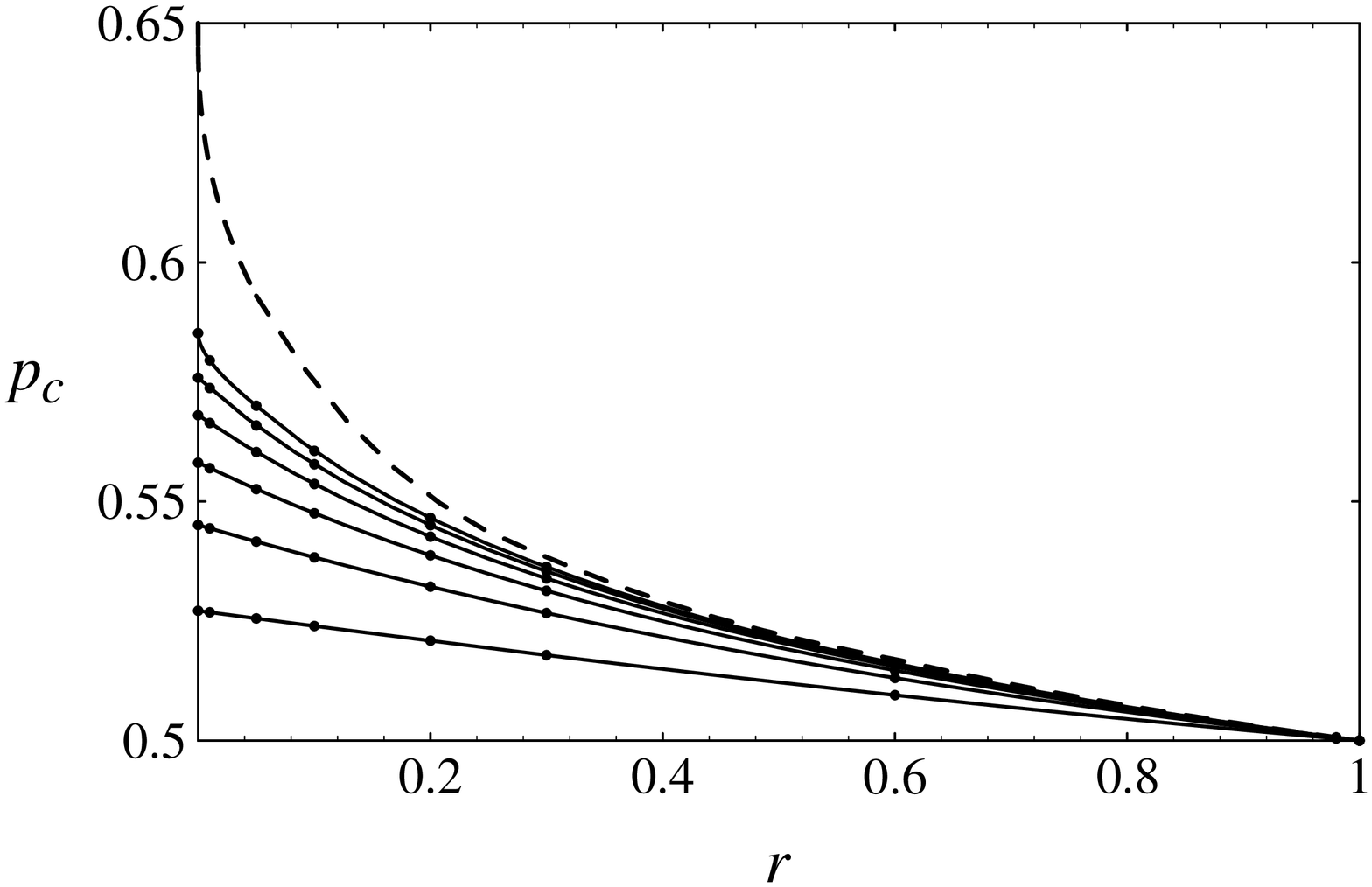}}}
\caption{Finite-size estimates for the percolation threshold
$p_c(r)$ from the locations of the minima in the energy mass gap.
Results for system sizes $N = 3, \ldots, 8$ (solid lines from bottom to 
top) are extrapolated to infinite $N$ (dashed line).}
\end{figure}

\begin{figure}
\centerline{\vbox{\epsfxsize=80mm \epsfbox{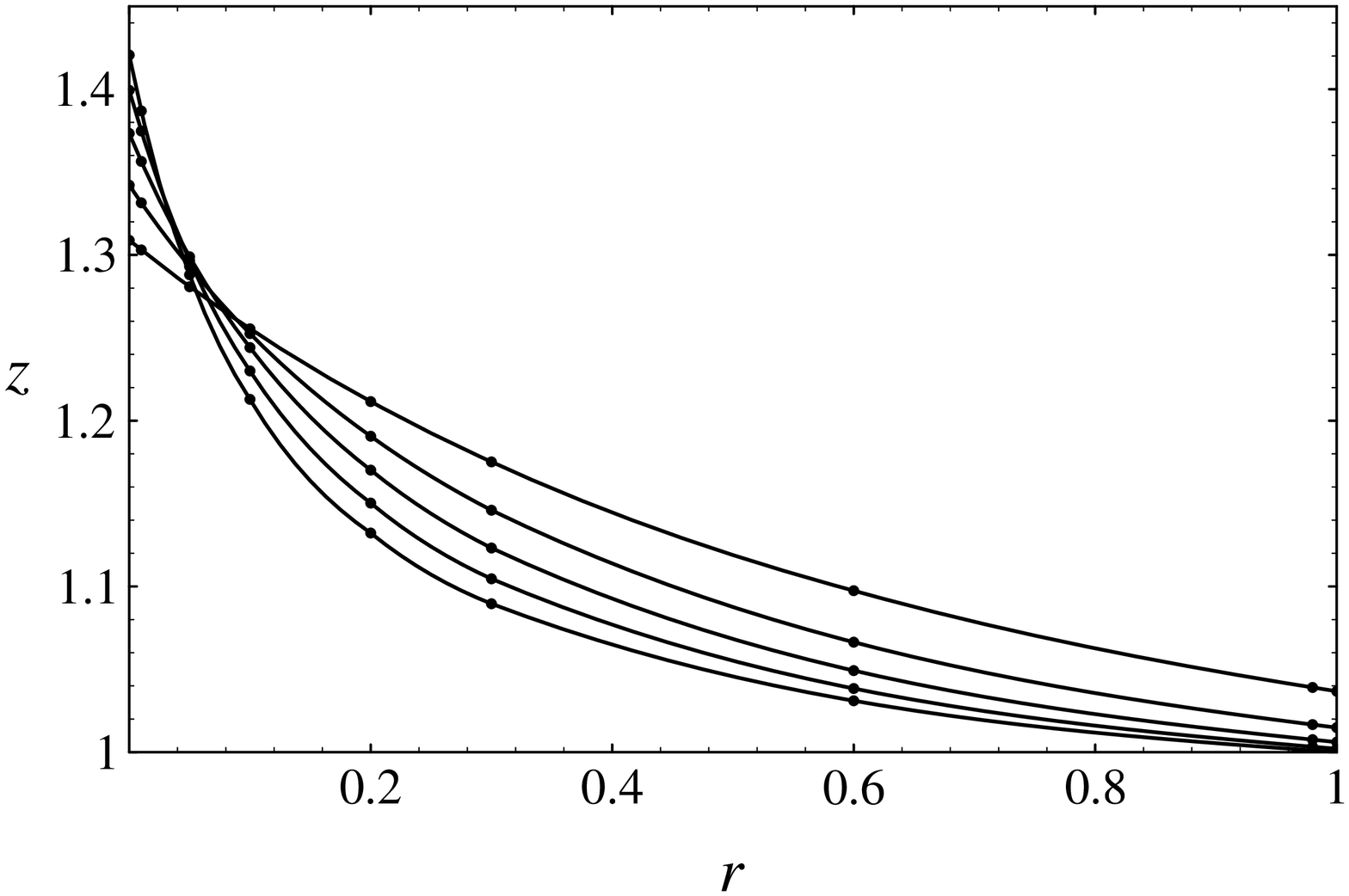}}}
\caption{Finite-size estimates $z(N,N+1)$ for the dynamic exponent 
$z$ using the relation $m \simeq A/N^z$. For $r>0$, $z(3,4), \ldots,
z(7,8)$ (from top to bottom) converge to $z=1$.}
\end{figure}

Fig.~2 shows FSS estimates for the location of the percolation threshold
$p_c(r)$ for system sizes $3\leq N\leq 8$.
The leading and next-leading eigenvalues of the transfer matrix
give us the energy mass gap, $m_e= \xi_\parallel^{-1}$,
the inverse of the time-like correlation length of 
energy-energy correlations.  
This mass gap is finite at either side of $p_c$. 
The minima in $m_e$  with respect to $p$ at each $r$
shown in Fig.~2 converge to $p_c(r)$.

Fig.~3 shows FSS estimates for the dynamic exponent $z$.
At $p_c$ the energy mass gap scales with system size as $m_e\simeq A/N^z$.
The values of the mass gap at the minima shown in Fig.~2,
yield for each set of system sizes (N,N+1) a FSS estimate for $z$.
These estimates converge to the IP value $z=1$ 
for all $r$, except very close to $r=0$.
Crossover scaling toward the DP value $z=1.581$ 
does not set in until $r$ is rather small.
FSS estimates for $y_p = 1 / \nu_\perp$ (not shown here) can be
obtained from the derivatives of $m_e$ with respect to $p$ at the minima.
These behave similarly to the estimate for $z$ in Fig.~3 and
converge towards the IP value $y_p =y_T =3/4$ for all $r>0$.
Crossover scaling towards the DP value does not play a role until rather 
small values of $r$. These numerical results clearly confirm that for all 
$r>0$ the scaling properties are identical to those of IP.
 
\begin{figure}
\centerline{\vbox{\epsfxsize=80mm \epsfbox{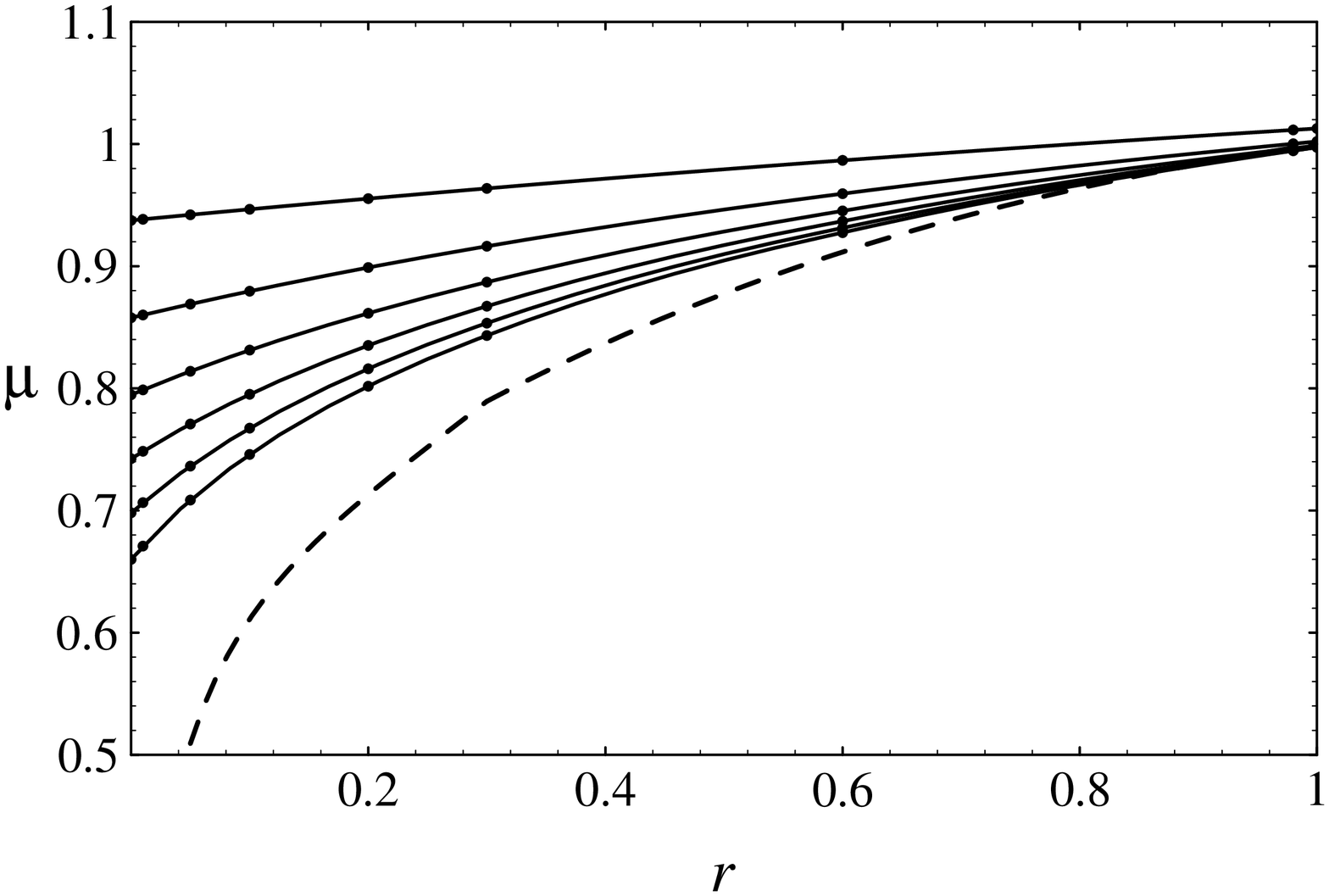}}}
\caption{Finite-size estimates for the anisotropy parameter
$\mu = \xi_\perp / \xi_\parallel$ at $p_c$ using the relation 
$m\simeq 2 \pi\mu (2-y_T)/N$ for system sizes $N=3, \ldots, 8$ 
(solid lines from top to bottom). The dashed line corresponds to the 
extrapolation $N \to \infty$.}
\end{figure}

Fig.~4 shows the FSS estimates for the aspect ratio 
$\mu=\xi_\perp/\xi_\parallel$ at $p_c(r)$. Both correlation lengths 
diverge with exponent $\nu=4/3$, but their amplitudes differ by a 
factor $\mu$. We obtain $\mu$ from the energy mass gap.
It scales according to conformal field theory as $m_e\simeq A/N$ 
with a universal amplitude $A=2\pi\mu (2-y_T)$.
In Fig.~4 we plot $N m_e / 2\pi(2-y_T)$ (with $y_T=3/4$) 
at the minima shown in Fig.~2.
For DP this ratio represents the so-called opening angle \cite{Kinzel}
and goes to zero at $p_c$.
The convergence to a non-zero value for $r>0$ gives further confirmation 
that the scaling is isotropic, $z=1$.

It is a safe guess that Eq.~(\ref{Cross_op}) and $y_{DP}=y_T-1$ 
hold also in dimensions larger than $D = 2$.
Eq.~(\ref{DP_op}) is easily generalized to higher dimensions.
The only aspect missing for general $D$ is the Coulomb-gas
rigorous proof that the scaling dimension
of the gradient of the energy maintains its naive value $y_T-1$.
The relation $y_{DP}=y_T-1$ suggests that IP changes stability 
with dimensionality.
The critical exponents for IP in 3D are known numerically \cite{3DPerc};
$y_T= 1.14$ yields $y_{DP} = 0.14$.
So for $D \geq 3$ the crossover operator appears to be relevant.
Mean-field theory, valid in $D \geq 6$, 
predicts that the crossover exponent is relevant as well. 
Power counting yields the value $y_{DP} = 1$.

Our identification of the crossover scaling field with the energy 
gradient agrees qualitatively with a recent field-theoretical analysis 
by Frey {\it et al.} \cite{FTS} (see also Ref.\ \cite{Luzhkov}).
They describe IP by a $\phi^3$-type theory similar to the 
conventional one for the $q$-state Potts model, and set up an 
interpolation scheme between this and Reggeon field theory \cite{Reggeon},
which belongs to the DP universality class.
Frey {\it et al.} perform a one-loop expansion about the upper critical 
dimension $D_u=6$ for IP and observe indeed a flow from isotropic to
directed percolation in their field theory.
Our approach has the advantage of being 
based on a well-defined microscopic model, and  being exact in 2D.

The apparent turn-about in the stability of IP for $2<D<3$ is the most 
intriguing implication of Eq.~(\ref{Cross_op}). 
Initially we thought that IP should be unstable towards DP in all dimensions. 
$r<1$ introduces a preferred direction in the model (the time-like direction). 
This breaks a fundamental symmetry of the problem and therefore it seems
logical that the scaling properties change. This is too simplistic, however. 
The crossover operator is the gradient of the energy density. 
$y_T$ is the fractal dimension of so-called ``red vertices'' \cite{IPrev},
i.e., of vertices where the cluster is singly connected (like the $\star$ 
states in our transfer matrix). Such vertices are rare, to the extent that 
in $D=2$ their fractal dimension becomes smaller than one, $y_T=3/4$.
This causes $y_{DP}$ to become negative.
  
The following argument puts the shoe on the other foot.
It supports the opposite point of view,
namely that DP at intermediate values of $r$ belongs to the
same universality class as IP (at $r=1$) in all dimensions.
Consider DP at $r=0$ just below its  percolation threshold.
All forest fires that are initially burning die out.
They are finite in size with a characteristic aspect ratio 
$\mu_0=\ell_\perp / \ell_\parallel$.
At small non-zero $r$, new forest fires are starting at all times $t$.
They have the same typical size and aspect ratio as those at $r=0$.
The probability that a specific vertex belongs to a forest fire
is of order $\tilde p\simeq r \ell_\perp^{D-1} \ell_\parallel$.
These forest fires are placed at random, and increasing
$r$ is equivalent to increasing $\tilde p$ in a conventional 
percolation problem, one in which objects of a specific size and 
shape are placed at random in space.
Let $\tilde p_c$ be the percolation threshold for that percolation problem.
The scaling properties at $\tilde p_c$ are identical to those at $r=1$.

The above argument suggests that DP is always unstable with 
respect to the creation of spontaneous forest fires.
It is probably too simplistic to treat $\tilde p_c$ as a constant,
but doing so leads to the following
estimate for the crossover exponent $y_{IP}$ at $r=0$.
The IP critical line approaches the DP critical point at $r=0$ as 
$\tilde p_c\simeq r \ell_\perp^{D-1} \ell_\parallel 
\sim r (p-p_c)^{-(z+D-1)\nu_\perp}$
with $p_c$ the DP percolation threshold at $r=0$.
According to conventional scaling theory the critical line must
approach the DP critical point along a renormalization flow line.
This yields $y_{IP}=z+D-1$ since $y_p = 1 / \nu_\perp$. 
This value can only be an approximation, since
it violates the fundamental rule that critical dimensions, like $y_{IP}$, 
cannot be larger than the embedded dimension. It shows however that 
DP is strongly unstable with respect to $r$ for all $D$.

This leaves us with a puzzle.
Do the scaling properties at intermediate values of $r$ 
change between $D=2$ and $D=3$, as suggested by the stability analysis at
$r=1$, Eq.~(\ref{Cross_op}), or not, as suggested by the above argument 
close to $r=0$?
Future numerical studies in $D=3$ will decide this issue.
For the time being, we put our bets on the above argument.
It is hard to see how it can be circumvented.
Moreover, the gradient nature of the crossover operator at $r=1$ is
suspicious. 
Gradient operators can be integrated up into boundary operators
and then often vanish from the theory altogether. 
This would leave the crossover at $r=1$ irrelevant in all $D$.

This research is supported by NSF grant DMR-9205125. 



\end{multicols}
\end{document}